\documentclass{vgtc}                          

\usepackage{times}                     

\usepackage{enumitem}
\usepackage[svgnames]{xcolor}
\usepackage{tabularx}
\usepackage{booktabs}
\usepackage{capt-of} 
\usepackage{multirow}
\usepackage{tikz}
\usepackage{wrapfig} 

\definecolor{CBgreen}{RGB}{102,194,165}
\definecolor{CBorange}{RGB}{252,141,98}
\definecolor{CBblue}{RGB}{141,160,203}

\definecolor{NavyBlue}{RGB}{0, 0, 128}

\usepackage{mathptmx}                  

\onlineid{1189}

\vgtccategory{Research}

\vgtcinsertpkg

\title{Towards Understanding Decision Problems\\As a Goal of Visualization Design}

\author{Lena Cibulski\thanks{e-mail: lena.cibulski@uni-rostock.de}\\ %
        \scriptsize University of Rostock, Germany %
\and Stefan Bruckner\thanks{e-mail: stefan.bruckner@uni-rostock.de}\\ %
     \scriptsize University of Rostock, Germany %
}

\def\arrow#1{%
\begin{tikzpicture}[
	node/.style={}
]
\draw[->] (0,-0.15) to (2.5,-0.15);
\node[node, align=center] at (1.25,0) {#1};
\end{tikzpicture}}
\teaser{
  \centering
		\small
		\captionof{table}{To concretize decision-support claims and inform downstream visualization design, we propose a framework of 15 situational variables that help describe key properties and conditions of a decision problem that are relevant for visualization support.}
		\begin{tabularx}{\linewidth}{clll}
			\toprule
			& Type & Leading Question & Spectrum \arrow{More challenging} \\
			\midrule
			\textcolor{CBgreen}{\textbf{Complexity}} & \# Options & How many options to choose from? & few vs. many \\
			\multirow{5}*{\includegraphics[width=0.06\textwidth]{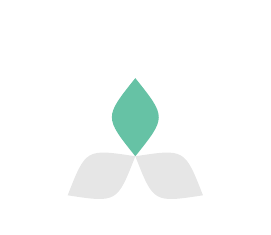} \includegraphics[width=0.06\textwidth]{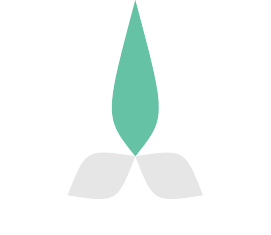}} & \# Attributes & How many criteria (attributes) to consider? & few vs. multiple \\
			& Availability & Alternatives generated before or during the decision? & known vs. progressive \\
			& Properties & All alternatives defined across the same attribute set? & comparable vs. non-comparable \\
			& Consequences & Potential outcomes known at the time of decision? & certain vs. uncertain \\
			& Preferences & Decision-maker can specify what is important? & clear vs. vague \\
			\midrule
			\textcolor{CBblue}{\textbf{Competence}} & Expertise & Decision-maker trained to make this kind of decision? & casual vs. professional \\
			\multirow{3}*{\includegraphics[width=0.06\textwidth]{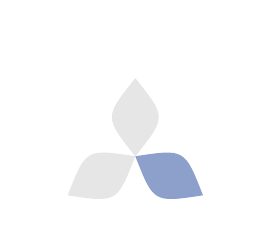} \includegraphics[width=0.06\textwidth]{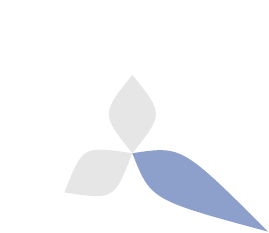}} & Authority & On whose behalf is the decision made? & oneself vs. others \\
			& Social setting & How many stakeholders are involved in the decision? & individual vs. group \\
			& Involvement & How much thought and knowledge go into trade-offs? &  low vs. high \\
			\midrule
			\textcolor{CBorange}{\textbf{Criticality}} & Stakes & How severe are the decision's consequences? & low vs. high \\
			\multirow{4}*{\includegraphics[width=0.06\textwidth]{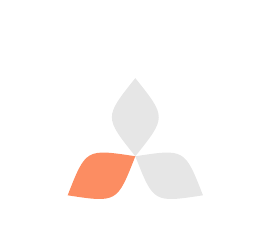} \includegraphics[width=0.06\textwidth]{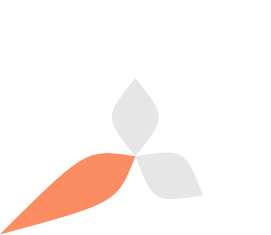}} & Time frame & How urgent is a decision? & not critical vs. critical \\
			& Frequency & How often and regularly is the decision made? & routine vs. ad hoc \\
			& Procedure & Is there an imposed decision strategy? & prescriptive vs. exploratory \\
			& Investment & When is the decision expected to pay off? & short-term vs. long-term \\
			\bottomrule
		\end{tabularx}
\label{tab:targeted-task}
}

\abstract{
    Decision-making is a central yet under-defined goal in visualization research. While existing task models address decision processes, they often neglect the conditions framing a decision. To better support decision-making tasks, we propose a characterization scheme that describes decision problems through key properties of the data, users, and task context. This scheme helps visualization researchers specify decision-support claims more precisely and informs the design of appropriate visual encodings and interactions. We demonstrate the utility of our approach by applying it to characterize decision tasks targeted by existing design studies, highlighting opportunities for future research in decision-centric visualization.
} 


\begin{document}



\maketitle

\section{Introduction}

Although decision-making is a long-standing goal in visualization research \cite{Bertin:BOOK83}, it
remains underexplored and ambiguously defined. Few works have studied the interplay of visualization and decision activities, e.g., how humans make decisions with visualized data or how well a visualization assists a real-world decision.
Dimara and Stasko attribute this under-representation of decision tasks, among others, to a lack of task clarity \cite{DS21}, as most decisions are ill-defined in the sense that they do not come with a formal definition of correctness and thus have no definite optimal solution \cite{RW72}.
To develop and evaluate visualization support, we need to clearly understand and describe the decision problem to design for as it is critical for all subsequent stages of the visualization design process \cite{Munzner09}. However, little guidance is available on how to concretize decision problems for visualization design.

Existing abstractions of visualization tasks \cite{BM13}, cognitive biases \cite{DFPBD18}, and knowledge generation models \cite{SSSKEK14} connect to decision-making as a high-level process, but do not consider the decision itself.
Disciplines like psychology and cognitive science summarize the foundations of how humans make decisions in process models \cite{Simon:BOOK60} or decision strategies \cite{PPBJ:BOOK93}, providing methods to describe the underlying intangible mental processes \cite{HCS98}, but often overlook the challenges posed by data-intensive decision environments typical for visualization. Visualization works follow these contributions to increase the decision focus of task analyses that inform visualization design \cite{DS21, TSM15, CDHK22}, but the properties of decision problems have not yet been investigated and described systematically.

Prior efforts to describe decision tasks, whether in reviews \cite{TSM15, DS21, BMAPC24} or design studies \cite{CMMK20, ZWCDCWLQ17, GLGPS13}, have focused on decision processes, i.e., \emph{how} a decision is made, rather than examining  \emph{what} factors and conditions shape a decision situation. Yet, the decision setting equally influences choices made during visualization design and evaluation.
One exception is a task analysis that uses situational features to identify three classes of decision scenarios and derive visualization recommendations on this basis \cite{HJCM20}.
However, this work targets group decision-making and the classification mainly serves as additional context for describing visualization tasks. Dimara et al. more generally refer to decisions as the task of "\emph{finding the best alternative among a finite set of alternatives, where alternatives are defined across several attributes}" \cite{DBD17}.
Although their definition accounts for certain decision properties, such as whether alternatives are known or whether a decision procedure is prescribed, additional properties are needed to accurately position a decision within the design space of decision problems.
However, in their survey of 27 decision-focused visualization tools, Oral et al. noted \emph{"difficulties in extracting concrete information about the decision elements, such as alternatives, criteria, and preferences"} \cite{OCWMD23}.

In this work, we take a first step towards filling this gap. We propose a characterization scheme that captures key properties of decision problems relevant for visualization.
Our framework organizes situational variables along three dimensions: the complexity of the decision space, the competence and autonomy of decision-makers, and the criticality of the decision context. By making these properties explicit, we aim to support researchers in (1) specifying the decision problems their visualizations address, (2) selecting appropriate design strategies, and (3) identifying promising directions for future work in decision-centric visualization.

\section{The Constructive Nature of Decision Tasks}
\label{sec:wicked-nature}

Decision-making is a multifaceted activity that has been examined from various disciplinary perspectives.
Many real-world decisions share a crucial trait: they lack a single optimal solution.
Instead, options differ in how well they satisfy multiple, often conflicting criteria, and it is up to the decision-maker to weigh these trade-offs.
Two perspectives emphasize this subjectivity \cite{DT03}: the \emph{prescriptive} view assumes that decision-makers retrieve pre-existing preferences, while the \emph{constructive} view holds that preferences are shaped during the act of decision-making. 

In this work, we adopt the constructive perspective \cite{Tsoukias08}, which aligns with the notion of \emph{wicked} \cite{RW72} problems. 
Both center around the idea that the problem and its solutions co-evolve.
In other words, decision problems are not fully defined from the outset. 
They unfold as information is structured, explored, and interpreted -- particularly when decisions are difficult or the decision-maker is inexperienced \cite{BZ77}.
Preferences are not fixed but constructed from beliefs, experiences, heuristics, and trade-off rules \cite{BZ77}.
Learning plays a central role in this process, connecting to theories of adaptive decision strategies \cite{PPBJ:BOOK93} and the way new insights reshape criteria and alternatives during exploration \cite{BLP98}.

To further illustrate the complexity of constructive decisions, we draw on four of Rittel and Webber’s characteristics of wicked problems that are particularly relevant to decision-making \cite{RW72}:

    \noindent\textbf{No enumerable set of potential solutions}\quad  Decision problems rarely involve a small, fixed set of options. Advances in computing have made it easy to generate large, nuanced alternative spaces, where each option differs across multiple dimensions.
    Knowing when to stop expanding the solution space becomes a matter of judgment. Decision-makers must sift through large volumes of data, often facing the challenges of scale and complexity.
    
    \noindent\textbf{No true-or-false solutions}\quad There is usually no objectively correct choice. While some options may be more desirable than others, decisions are shaped by subjective preference, intuition, and expertise. These vary between individuals and may evolve during the process as decision-makers adjust to what is realistically achievable.
    
    \noindent\textbf{No stopping rules}\quad Without a clear benchmark for success, it is difficult to know when to stop. New options may always emerge, and improving one criterion often worsens another. For example, discovering better warranties may justify increasing a car budget. Desired outcomes can be fluid, requiring ongoing trade-offs. 
    
    \noindent\textbf{No transferability}\quad Even familiar decisions do not repeat exactly. Each one is shaped by changing technologies, goals, and contexts. Replacing an old camera, for example, involves a new market landscape and possibly different needs. Prior experience may help, but each situation requires fresh evaluation.

In summary, decision problems are inherently dynamic: available information, user goals, and evaluation criteria evolve during the decision process. We aim to bridge the gap between the prevalence of multi-attribute decision tasks and the limited attention visualization research has given to decision-specific information needs.

\section{Situational Analysis of Decision Problems}

While each decision problem presents unique challenges, we perform a situational analysis to abstract the conditions that frame a decision to inform the design of appropriate visualization support.
Through constant literature review and consultation of domain experts in the authors' prior research on visualization-based decision support \cite{Cibulski24}, we identified common properties across a broad range of decision contexts. 
Building on previous work on \emph{situational variables} \cite{HJCM20}, we propose a systematic description of the factors that influence how people approach decisions with visualizations.
Our goal is to encourage a more rigorous focus on decision-making as a core objective of visualization research. 
While our proposed list of situational variables aims to cover the design space of decision problems, it is by no means immutable.
As design experience and task clarity evolve, it will likely be subject to refinement. 

In a bottom-up approach, we identified 15 situational variables,  organized into three overarching dimensions of a decision problem:
\begin{itemize}[noitemsep,label={},leftmargin=0pt, topsep=4pt]
    \item \textcolor{CBgreen}{\textbf{Complexity}}: How complex is the decision space, i.e., what is the nature of the decision elements such as alternatives or criteria?
    \item \textcolor{CBblue}{\textbf{Competence}}: What expertise, autonomy, and stakeholder structure do decision-makers have?
    \item \textcolor{CBorange}{\textbf{Criticality}}: How severe and constrained is the decision context?
\end{itemize}

For each situational variable, we formulated a leading question and defined a continuum of possible levels, from a level that does not complicate decision-making to a level that significantly complicates decision-making (e.g., not time-critical to highly time-critical).
The results are shown in Table \ref{tab:targeted-task}.

A complicating factor in one of the situational variables poses a complication for the respective upstream dimension (i.e., complexity, competence, or criticality).
In turn, each dimension can pose a low or high complication to the decision problem depending on different manifestations in its associated situational variables.
If we allow each dimension to take one of the values "low" or "high" complication, we have eight possible combinations, each of which describes a class, or scenario, of decision problems.
For a better recall, we represent each scenario by a flower with three petals $\vcenter{\hbox{\includegraphics[scale=.1]{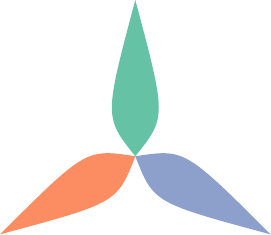}}}$ representing each of the three dimensions \textcolor{CBgreen}{complexity}, \textcolor{CBblue}{competence}, and \textcolor{CBorange}{criticality}. 
The length of a petal indicates low or high complication for that dimension (cf. first column in Table \ref{tab:targeted-task}).

\noindent \textcolor{CBgreen}{\textbf{Complexity}}: The complexity dimension refers to the decision space and involves characteristics of the options, criteria, and preferences. Decisions are relatively simple when options are few and trade-offs are clear. They become more complex if the number of alternatives or criteria gets larger or preferences are vague. 
The challenge also increases if decision options are not comparable, generated on the fly, or their outcomes are uncertain.
Additionally, some decisions involve complex underlying phenomena that must be understood before meaningful choices can be made.

\noindent \textcolor{CBblue}{\textbf{Competence}}: Competence captures the decision-maker's expertise, autonomy, and the social setting of the decision. 
Decisions in a personal context are likely easier to make than professional decisions that require consensus among multiple parties and/or whose consequences affect large groups of people.
Furthermore, decisions become more challenging when they require extensive exploration of trade-offs, rather than relying on recommendations grounded in well-defined goals or rules.

\noindent \textcolor{CBorange}{\textbf{Criticality}}: The criticality dimension refers to the severity and constraints that frame a decision to be made.
Decisions that arise periodically, do not have severe consequences, or can be reverted with low effort are relatively easy to make and justify.
In contrast, they become more challenging when they occur ad hoc, when they need to be made under time pressure, or -- most critically -- when human health or even survival depends on them. 

\section{Decision Scenarios Spanning the Design Space}

To illustrate how the complexity, competence, and criticality dimensions combine in practice, we outline eight representative decision scenarios. These scenarios are derived from all possible combinations of low and high complication levels in each dimension. 

\begin{wrapfigure}[3]{l}{0.02\textwidth}
\vspace{-12pt}
\includegraphics[width=.06\textwidth]{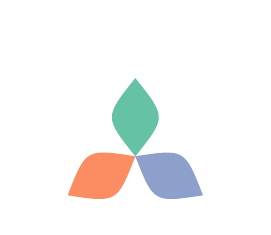}
\end{wrapfigure}
\noindent \textbf{Micro Decisions} Micro decisions involve a simple, manageable decision space and low stakes, requiring no specialized skills.
These are typically personal everyday choices that do not involve a lot of options or critical conditions.
They can range from binary decisions, e.g., whether to have coffee or tea, to multi-attribute decisions, e.g., which film to watch.
They also involve routine purchases or rentals, e.g., which book to borrow from a library, where a bad decision can be easily reverted.
Such familiar decisions are often informed by the decision-maker's current sentiment rather than fully developed preferences.
Although visual data analysis does not commonly target this scenario, limited decision support is offered by online marketplaces, e.g., tables that compare a product of interest to few similar products across multiple criteria.
It is also supported by algorithmic recommenders that often surface
options without requiring explicit preference input.

\begin{wrapfigure}[3]{l}{0.02\textwidth}
\vspace{-12pt}
\includegraphics[width=.06\textwidth]{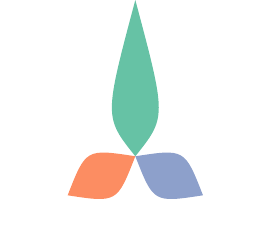}
\end{wrapfigure}
\noindent \textbf{Hobbyist Decisions} Hobbyist decisions involve higher complexity, but are uncritical and can still be made without professional skills.
A typical example are decisions that are made for one's own enjoyment, e.g., related to hobbies, or based on personal data, i.e., data about oneself or relevant to personal interests.
They might involve complex options but do not involve severe consequences, even in the worst case. 
A lot of thought can go, for example, into deciding which plants to place where for gardening, which facilities to develop for civilization-building strategy games, how to name a newborn \cite{Wattenberg05}, or how to balance ingredients for a healthy nutrition.
For example, \emph{Dust\&Magnet} allows to choose cereals based on their dietary composition by attracting the diverse options with attribute anchors in a 2D projection \cite{YMSJ05}.
Pousman et al. reflected on visual representations that depict data in everyday situations and might be viewed only during a fleeting moment \cite{PSM07}.
Huang et al. have surveyed the design space of personal visualizations, which help individuals gain insights about themselves and their interests \cite{HTAB+14}.
Another example are consumer choices from a catalog, which might involve many options and criteria to consider but typically do not require specialized knowledge, e.g., which of the hundreds of hotels in a city to book for a vacation.
For example, the decision support tool \emph{SkyLens} was used to decide for a city to visit for a one-month holiday \cite{ZWCDCWLQ17}.

\begin{wrapfigure}[3]{l}{0.02\textwidth}
\vspace{-12pt}
\includegraphics[width=.06\textwidth]{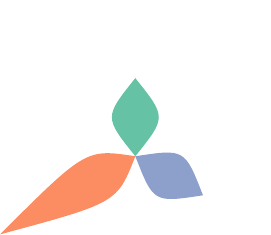}
\end{wrapfigure}
\noindent \textbf{Emergency Layperson Decisions} Emergency layperson decisions are decisions that are not per se difficult for laypersons to make, but might involve severe consequences in case of an inappropriate decision.
Criticality might emerge from time pressure, e.g., when laypersons need to react as emergency-responders. 
Deciding about how to appropriately respond in case of, e.g., fire, is critical, but does not require specialized knowledge or in-depth investigation of a variety of alternatives or decision criteria.
While the development of solutions for these types of decision has so far not been a focus of visualization research, studies have investigate the interpretation of visual representations, e.g., to estimate hurricane risk \cite{LBRP+16} or for deciding to leave one's home in case of wildfire \cite{MHTD24}.

\begin{wrapfigure}[3]{l}{0.02\textwidth}
\vspace{-12pt}
\includegraphics[width=.06\textwidth]{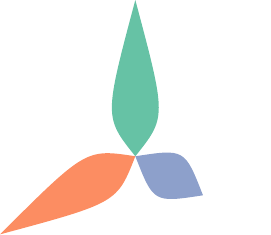}
\end{wrapfigure}
\noindent \textbf{Personal Life Decisions} Personal life decisions are typically complicated and might come with extensive research, but do not require year-long training or education.
This is due to life choices typically being non-routine, ad hoc decisions that pose individual cases, such as buying a house, choosing a university to attend \cite{GLGPS13}, or job seeking.
They are critical in that they have a lasting effect on a person's life.
Still, they are typically made on behalf of oneself and do not require justification beyond that.
Complexity can originate from a large number of options or from criteria being difficult to assess and/or conflicting, e.g., when comparing the services included in an insurance offer, balancing risk and return of stocks to buy, or deciding about individual pension provision solutions.
Another representative example are laypeople serving as jury members at a court, where they decide whether a person should go to jail without a professional education. 
In contrast to hobbyist decisions, life decisions can thus have significantly more severe consequences.
They can be complicated if the expected consequences are subject to uncertainty, e.g., the reliability of a newly released car or the quality of a home's construction, which may only become apparent years after purchase. Visualization research supports, e.g., the task of home finding:
\emph{ReACH} particularly focuses on selecting the best location based on spatial constraints representing daily routines \cite{WZBZW18}, whereas \emph{RankASco} creates a ranking of suitable apartments based on flexible user-specified attribute preferences \cite{SCHB22}.
\emph{FinVis} is a visual analytics tool that helps non-experts interpret the return, risk, and correlation aspects of portfolio options to facilitate their financial planning \cite{RSE09}.

\begin{wrapfigure}[3]{l}{0.02\textwidth}
\vspace{-12pt}
\includegraphics[width=.06\textwidth]{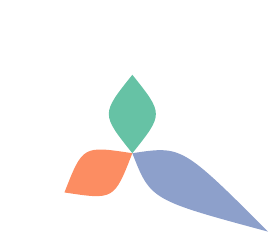}
\end{wrapfigure}
\noindent \textbf{Professional Routine Decisions} Professional routine decisions are generally not complicated or critical as they are made repeatedly under similar conditions.
Still, a professional background is needed to make an appropriate choice.
This can be business decisions, e.g., professional buyers placing orders for their company.
Another example is a group of senior researchers responsible for choosing an award recipient.
Tools like \emph{SkyLens} \cite{ZWCDCWLQ17} or \emph{Podium} \cite{WDCKBE18} might be used by sports journalists to decide on outstanding college football teams to write about or by basketball teams to decide which player to sign.
Visualization research has also provided solutions to help marketing analysts advertise services through e-mail \cite{GDMK+19}, regulatory inspectors decide whether reported fishing activities are legal \cite{SB20}, meteorologists broadcast accurate short-term weather forecasts \cite{DPDS+15}, or credit analysts make a fair choice about the customer to grant a loan \cite{AL19}.
\emph{bCisive} has been designed to streamline decision-making in a business context \cite{MSGPB10}.
Still, professional routine decisions are not at the core of visualization research, since commercial systems are typically available.

\begin{wrapfigure}[3]{l}{0.02\textwidth}
\vspace{-12pt}
\includegraphics[width=.06\textwidth]{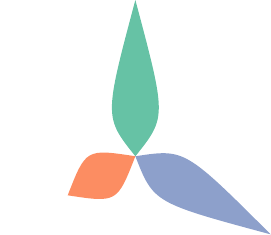}
\end{wrapfigure}
\noindent \textbf{Optimization Decisions} Optimization decisions are informed by a complex information basis and require professional expertise. 
While their consequences should be taken seriously, they are usually not critical in a "pass-or-fail" sense.
In most cases, the available time frame matches or even exceeds the time required to make such a decision. 
Expert decisions are typically made on behalf of other stakeholders.
Due to their nature, optimization decisions have long been subject to visualization design studies that are conducted in close collaboration with domain experts.
The underlying data can include attributes, space, and time as facets.
Solutions might follow different strategies like ranking or multivariate exploration.
Visualization tools can facilitate the specification of geometry, material, etc. such that an electric drive's properties optimize the customer's application-specific requirements \cite{CMSK22}.
Similarly, visualization has been used to identify the most preferred configuration for automatic transmission in a powertrain \cite{PSTSMP16}.
Solutions have also been proposed for trade-offs in fisheries management \cite{BMPM12}, deciding among different lighting setups for buildings \cite{SOLSGP15}, or deciding among software release plans \cite{AWLRTC15}. 
Further examples are urban planners tasked with the identification of suitable areas for new buildings \cite{FLDV+15} or the adjustment of bus route networks to travel demands \cite{WZDM+20}.
Optimization decisions are often made among a fixed set of solutions, e.g., from pre-computed simulations. 
However, new solutions in a region of interest might also be generated progressively during the decision process, e.g., an operator steering the optimization of wastewater treatment.

\begin{wrapfigure}[3]{l}{0.02\textwidth}
\vspace{-12pt}
\includegraphics[width=.06\textwidth]{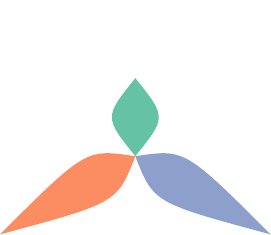}
\end{wrapfigure}
\noindent \textbf{Emergency Professional Decisions} Emergency professional decisions have similar characteristics as emergency layperson decisions.
The main difference is that professional education, training, or experience are needed to quickly identify and interpret those signs that inform the required course of action.
An example might be medical personnel tasked with the triage in an emergency room.
Many of these decisions need to be made under time pressure. 
For example, police dispatchers need to forward incoming emergency calls to the correct operational units.
Similar to this is the role of traffic controllers, e.g., in aviation, where visualization can support the detection and resolution of potential flight conflicts \cite{ANDC99}.
We could also think of a fireground commander being called to an emergency and having to decide whether to initiate search and rescue.
While requiring professional expertise, these choices usually involve only the knowledge, intuition, and judgment of the decision-maker, as opposed to a group decision setting.
Visualization tools have also been proposed in the field of emergency management to support the task of evacuation planning in a disaster-affected area \cite{RHDW12}.
To inform emergency response in the case of severe weather events, e.g., a tornado outbreak, \emph{Hornero} supports the detection and nowcasting of thunderstorms \cite{DPRP+21}.

\begin{wrapfigure}[3]{l}{0.02\textwidth}
\vspace{-12pt}
\includegraphics[width=.06\textwidth]{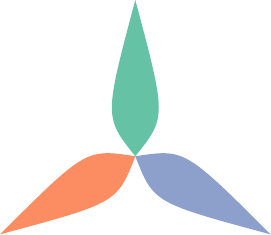}
\end{wrapfigure}
\noindent \textbf{High Stakes Decisions} High stakes decisions are extremely challenging to make due to their complexity, their criticality, and the in-depth professional knowledge that is required.
High stakes, in terms of affected health or even survival but also time pressure, are associated with many decisions in medical treatment planning, e.g., which patient should receive an organ.
Visual assistance tools support decisions in clinical environment based on, e.g., event sequences \cite{LLP*22} or blood flow data \cite{PGLMP14}.
High stakes decisions also occur in politics, e.g., when needing to agree on the appropriate reaction in an armed conflict that threatens complete populations.
Multidisciplinary tumor boards planning patient treatment \cite{MCOSZSWDO21} or politicians deciding about a measure to be implemented are collaborative decisions and are thus complicated by the need to hear all relevant stakeholders and arrive at a consensus.
In the field of public health, visualizations help explore epidemic models and their impact to inform decisions about effective measures in the case of potential mass casualty events like a disease outbreak \cite{AME11}.
Similarly, \emph{ManyPlans} helps generate and evaluate protective measures for different flooding scenarios \cite{WKSHRCKS14}.
Due to their criticality and the fact that high stakes decisions are usually made on behalf of others, they come with particular demands regarding transparency, justification, and documentation.

\section{Implications for Decision-Centric Visualization}

The eight decision scenarios outlined in our framework underscore the importance of tailoring visualization design to the specific context in which a decision occurs.
Here, we reflect on the three dimensions of our proposed design space more generically against the background of visualization research. 

\noindent \textcolor{CBgreen}{\textbf{Complexity}}: \emph{Higher complexity requires more advanced visual mappings and user guidance.} 
For low-complexity decisions, with few options and clear goals, simple visualizations such as basic tables or charts often suffice. In contrast, high-complexity tasks, where numerous options and criteria are involved, require more sophisticated visual techniques. For example, \emph{Podium} \cite{WDCKBE18} uses a familiar tabular view to convey multi-attribute rankings, while \emph{WeightLifter} employs multiple complex linked views for powertrain optimization \cite{PSTSMP16}.
To manage clutter in complex scenarios, advanced aggregation or focus-and-context techniques should be used.
For choices with disjoint attribute sets, hybrid visual encodings (e.g., parallel coordinates combined with heatmaps) may help. 
If alternatives are progressively generated, visual design should focus on an interaction with optimization algorithms rather than static alternatives.
Furthermore, when consequences of a particular alternative are uncertain, visualizations need to transparently communicate uncertainty information to avoid decision biases.
Likewise, when decision-makers lack established preferences, tools should help construct and refine preferences interactively (e.g., by visually comparing alternatives or by recommending sensible defaults).

\noindent \textcolor{CBblue}{\textbf{Competence}}: \emph{Higher competence requires increased support for specialist insight and analytical thought.}
Visualization design should align interaction complexity and control with the user’s level of expertise. For instance,  \emph{Dust \& Magnet} \cite{YMSJ05} enables casual users to explore multivariate data through an intuitive metaphor, while Weng et al. \cite{WZDM+20} offer specialized interfaces for transportation system analysts to incorporate their expert knowledge about potential causes for bus transport anomalies.
Decision-makers range from trained experts who make repeated decisions in their domain to novices who engage with data occasionally in personal contexts.
Understanding their motivation, priorities, environments, and cognitive processes is essential for deriving appropriate design requirements.
Factors such as personality traits, cognitive abilities, visualization literacy, and domain expertise have been shown to influence how users interact with visualizations \cite{CCHST14}.
For expert users, visualizations should offer rich, configurable representations that enable in-depth exploration, comparison, and interpretation. They benefit from advanced interaction techniques and mechanisms to capture and communicate decision rationale, such as annotations, provenance tracking, or explanation views, especially when decisions must be justified to others. In group settings, shared views or features for comparing stakeholder preferences can further support consensus building. In contrast, casual or novice decision-makers require highly usable, lightweight tools that minimize interaction overhead and guide interpretation. Simpler visual encodings and intuitive metaphors can reduce cognitive load and foster engagement. In such settings, supporting reflective insight rather than precise analysis may be more appropriate. As such decisions often involve personal taste or emotion, visual aesthetics and clarity play an important role in building trust and encouraging exploration.

\noindent \textcolor{CBorange}{\textbf{Criticality}}: \emph{Higher criticality calls for visualization designs that support thoughtful, reflective reasoning.}
When the stakes are low -- such as choosing a holiday destination -- simple, familiar visualizations often suffice and should emphasize usability over complexity. 
In contrast, high-stakes decisions, like those faced during a pandemic \cite{AME11}, demand tools that support deep understanding, clear communication, and accountability.  Visualizations in such contexts must make the underlying data, assumptions, and consequences transparent, helping users trace the basis for a decision and communicate it to others.
Provenance features, rationale capture, or decision summaries can support these needs.
For routine decisions made repeatedly with the same tool, interface familiarity can outweigh initial learnability, while ad hoc high-stakes decisions benefit from guided exploration and safeguards against misinterpretation. In time-critical situations, visualizations must prioritize speed and clarity, offering streamlined views that reduce cognitive load and support rapid yet informed responses under pressure. Aligning visual design with the urgency and consequences of a decision ensures that users can act both decisively and responsibly.

Overall, the scenario framework suggests that effective support tools should be tailored to the decision’s demands: complexity calls for features to manage many options or criteria, competence level dictates the necessary ease-of-use vs. advanced control, and criticality imposes requirements for clarity, robustness, and trust.

\section{Conclusion}

We introduced a framework for characterizing decision problems in visualization along three key dimensions: complexity, competence, and criticality. By combining these dimensions, we identified eight representative scenarios that reflect the diverse conditions under which people make decisions. Our analysis shows that each scenario places distinct demands on visualization design and evaluation, and that current research is unevenly distributed across this space. 
While the framework is intentionally descriptive, it offers a first step towards a structured vocabulary for reasoning about decision-making in visualization. 
Future refinement of the framework could benefit from more rigorous and structured methods to identify, confirm, and prioritize situational variables.
Likewise, future empirical evaluation is important to strengthen the grounding of our framework.
Finally, future decision support tools and design patterns whose development is informed by the framework can demonstrate its practical use and benefit.
We hope our work encourages visualization researchers to contribute to the evolution of this framework, to more precisely articulate which decisions their solutions aim to support, and to tailor their approaches accordingly.


\bibliographystyle{abbrv}

\renewcommand{\scriptsize}{\small}
\renewcommand{\baselinestretch}{0.96}

\bibliography{references}
\end{document}